Is jet re-orientation the elusive trigger for star formation suppression in radio galaxies?

David Garofalo[1], Emily Moravec[2], Duccio Macconi[3], Chandra B. Singh[4]

1. Department of Physics, Kennesaw State University, USA
2. Green Bank Observatory, USA
3. Department of Physics and Astronomy, Universita' di Bologna, Italy
4. South-Western Institute for Astronomy Research, Yunnan University, Kunming 650500, China; chandrasingh@ynu.edu.cn

*Abstract*

Jet re-orientation associated with the time evolution of radio quasars explains the formation of X-shaped radio galaxies and their preference for isolated environments. But since X-shaped radio galaxies are generally not found in dense environments (e.g. groups/clusters), the jet re-orientation phenomenon for radio galaxies in groups and clusters has been ignored. We take a closer look at the re-orientation of FRI jets with respect to FRII jets, and find that it may constitute the as-yet unidentified trigger for star formation suppression in radio galaxies. We show how the recently explored radio "red geyser" galaxies can be interpreted in this context and ultimately reveal a deeper understanding of why FRII radio galaxies are on one side of the star formation enhancement/suppression divide compared to FRI radio galaxies.

*Introduction*

Singh et al (2021) have shown that radio galaxies in different environments are distributed on the star formation rate – stellar mass plane (SFR-SM plane; Comerford et al. 2020) in a way that is compatible with the idea that Fanaroff-Riley II (FRII) quasars are counterrotating accreting black holes that evolve toward corotating black holes (Garofalo, Evans & Sambruna 2010). According to this picture, when a radio quasar is formed, it begins its life by moving upwards and to the right from the radio quiet line, as a result of the star formation enhancement associated with FRII jets (e.g. Capetti et al 2021; Kalfountzou et al 2014). This is in opposition to the decrease in star formation rate associated with a jetless, or radio quiet, AGN. In other words, while stars form at an increasingly lower rate in radio quiet AGN due to the effect of the disk wind on the ISM, radio loud AGN experience a phase in which the star formation rate increases for a few million years as the jet energy pushes the gas into a denser state. This phase is then followed by a star formation suppression phase that is stronger than that due to the disk wind, associated with corotation between black hole and accretion disk and an FRI jet. These two phases are shown for fields, groups, and cluster environments, in Figure 1. The change from FRII to FRI jets, which involves counterrotation evolving into corotation, implies a transition through zero black hole spin, the absence therefore of the Bardeen-Petterson effect, and a possible re-orientation of the accretion disk. If the accretion disk changes direction and a jet is formed, the jet will also experience a change in direction (Garofalo, Joshi et al. 2020). In short, radio galaxies that live on the green paths of Figure 1 may display jet orientation that is different from the jets of radio galaxies that live on the pink paths of Figure 1.

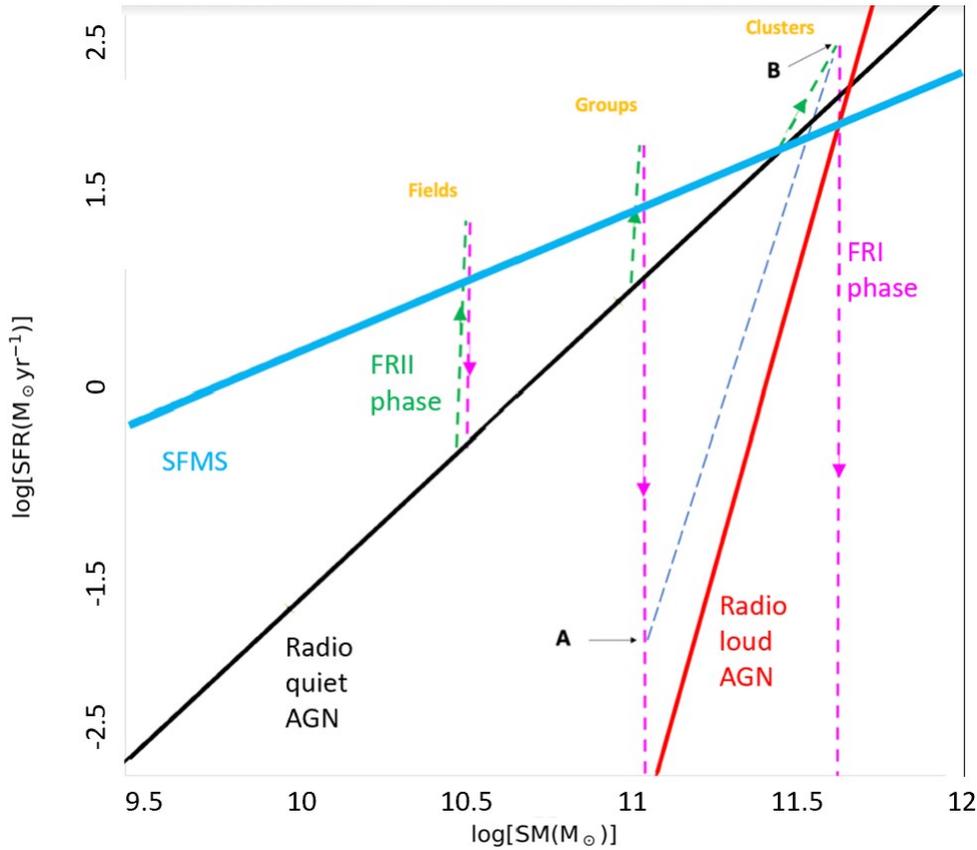

Figure 1: The predicted distribution of radio galaxies on the SFR-SM plane, modified Garofalo & Mountrichas (2022) version of figure from Singh et al (2021). FRII phases are green and evolve upwards while FRI phases are in pink and evolve downwards. Because of rapid evolution of thin disks to ADAF in cluster environments on average, the evolution in clusters for radio AGN is slower. When the average radio AGN in a cluster has reached the turning point at B between its FRII phase and its FRI phase, the average radio AGN in a less dense environment has already experienced an FRI jet for about a billion years which means it has also experienced a significant drop in its star formation rate, which places it at point A in the diagram.

Roy et al. (2021) have recently explored a subclass of 42 "red geyser" galaxies, a class of low redshift, low star forming galaxies, that were detected in radio surveys. They are low redshift (z < 0.1) and generally low star forming. While only a few red geysers have large scale jets, most displayed compact radio emission. For the three red geysers that have large scale radio emission, they find evidence of large-scale ionized winds whose directions appear to be either aligned or orthogonal to the radio emission. For the compact red geysers that appear as radio point sources, it is difficult to infer a direction for the radio jet, while it is unambiguous for the triple red geysers.

Compact red geyser galaxies may be giving us insight into the initial jet phase of a radio AGN. If the winds are due to the accretion disk (Kuncic & Bicknell 2007; Silk & Rees 1998) and the radio emission is associated with the jet (e.g. Padovani 2016), we are in a space of testable model predictions. According to the model presented in Singh et al (2021), in the early phases, the jet and winds are aligned (green paths), while in the later phases of its life, the new jet and wind directions may not be aligned. It is important to emphasize from the model perspective that we are comparing a newly formed jet associated with a spinning black hole that transitioned through zero spin (i.e. an older jet during sufficiently high spin in counterrotation and a newer jet associated with sufficiently high spin in corotation) with the wind that formed during the older jet phase and whose signature can be measured on large scales. Phases when this new jet and old wind are not aligned are predicted to be at least 10 million years after the radio source was triggered. We will show that the triple radio sources of Roy et al. (2021) that appear to display orthogonal directions between winds and jets, live further to the right and down on the SFR-SM plane, as expected in the model, and then proceed to generalize this behavior as a function of environment.

We also add a new feature to the model (the Roy Conjecture), that the re-orientation of jets constitutes an effective way to suppress star formation (idea emerged from a communication with Namrata Roy). We replace the previous idea/assumption that the morphology of FRI jets is responsible for more effectively distributing jet energy to the inter-stellar medium (ISM). The idea was based on the thought that the weaker collimation of FRI jets may allow the jet to interact more directly with its environment. With the Roy Conjecture, the effectiveness in shutting down star formation for FRI jets acquires greater depth, that spraying the galaxy with jets in a different direction more directly couples jet energy to the ISM. This suggests that collimation is not the key to affecting the ISM. The giant, so-called red-and-dead, elliptical galaxies in the densest cluster environments, will be understood to have evolved into a regime with low SFR due to the long-term effect of a re-oriented FRI radio jet, which effectively heats the ISM. This conjecture generates predictions across the active galaxy phenomenon and is ultimately likely to have impact in understanding the black hole scaling relations.

In Section 2 we describe the distribution of radio galaxies on the SFR-SM plane, place red geysers on that plane in Section 3 and make predictions, then conclude in Section 4.

*2. The time evolution of radio galaxies as a function of environment*

Singh et al. (2021) explored the connection between black hole and star formation as a function of environment. Star formation processes are enhanced in some active galactic nuclei (AGN) while suppressed in others. In the case of the minority population of jetted (radio mode) AGN, there are two phases wherein star formation is first enhanced followed by a longer phase in which it is suppressed. In the case of isolated field environments with the least massive black holes on average[1], the initial FRII phase is followed by a relatively shorter corotation phase (compared to corotation phases in denser environments), during which both jets and disks suppress the SFR (note the short pink path for fields in Figure 1). As one explores the formation of FRII jets (and thus counterrotating disks) in denser environments, one finds that jets in some sense win the tug-of-war they engage in with the accretion disk, and therefore manage to alter the state of accretion from a thin disk, to an advection dominated (ADAF) flow (Antonuccio-Delogu & Silk 2010). If the FRII jet fails to generate this change in the accretion flow, the disk persists in a radiatively efficient state which eventually smothers the jet in what is referred to as "jet suppression" (Neilsen & Lee 2009; Ponti et al 2012). The reason for this is that as the spin increases into the high corotation phase, the inner edge of the disk moves closer to the black hole and this allows the disk to tap into the gravitational potential energy close to the black hole. As a result, the radiative wind formed in the disk will dominate over the loading of disk gas onto magnetic field lines, quenching or suppressing the jet (Garofalo, Evans & Sambruna 2010). The type of feedback described above in which the heat transfer to the accretion flow to make it an ADAF, allows corotation and FRI jets to become the longest lasting combination of features of any AGN subclass.

Because the transition into the corotating phase for radio sources in more isolated environments tends to be accompanied by a smaller decrease in the excitation level compared to the average radio galaxy that transitions into the corotating phase in denser environments, such objects on average tend to be subject to jet suppression. This is the above-mentioned phenomenon of the disappearance of the jet for sufficiently high spinning corotating systems. As a result, such objects exit the radio galaxy classification which explains the shorter pink phase for field objects in Figure 1. The larger decrease in excitation level experienced in denser environments compared to isolated ones, by contrast, means that disks experience a wind suppression but a lack of jet suppression. Because the transition toward low excitation occurs more rapidly on average in the richest environments, the time evolution during the FRII SFR enhancement phase for clusters, is longest. Such systems experience both the longest SFR enhancement as well as the longest SFR suppression (see the greater tilt in the green path for clusters in Figure 1). While counterrotating (FRII phases) generate both disk winds and jets, we see that isolated and rich environments on average tend to experience an absence of jets and disk winds, respectively. As the counterrotating phase comes to a close as the black hole spin approaches zero spin, the jet turns off but the disk wind persists even through zero spin. At zero spin a new disk plane is possible due to absence of the Bardeen-Petterson effect, and a new jet will form once the spin is sufficiently high. Our goal is to compare the directions of this new jet with the direction of the large-scale wind produced by the accretion disk during the counterrotating phase. These two directions need not be the same. Given these ideas, in the next section we interpret the red geysers depending on their location on the SFR-SM plane.

---

[1] A rough range for black hole masses in isolated, intermediate, and dense environments is order $10^6$-$10^8$, order $10^7$-$10^9$, and order $10^8$-$10^{10}$ solar masses, respectively.

*3. Red geysers*

Here we determine the location of red geysers on the SFR-SM plane and connect them with the ideas presented in Section 2. We took the sSFR from Roy et al (2021) and multiplied by the total stellar mass to determine the SFR. The basic idea from theory is that radio galaxies that occupy more leftward and upward locations in Figure 1 (Compact sources) will tend to have jets and winds that are aligned while the opposite will be true for more rightward and downward sources. The compact red geysers display point radio emission, possibly making them final or initial jet phases. Such phases are predicted to be located near the turning points of Figure 1, close to the intersection between green and pink paths. They should, thus, tend to distribute themselves further left and up compared to late phases. Because compact red geysers are point radio sources, the direction of the radio jet cannot be determined and alignment between jet and wind cannot be determined. Nonetheless, we can predict that such objects will tend to be mixed aligned and non-aligned. The aligned sources can be either on green paths or pink paths. The non-aligned sources would have to be pink paths. This is because in the transition through zero black hole spin, the orientation of the accreting material may or may not change. We plot compact red geysers as squares showing that possible end or beginning jet phases are located in regions where star formation has yet to be strongly suppressed, as expected theoretically.

The remaining Irregular, Triple, and Double red geysers that we plot have jet directions that can be inferred and compared to wind direction. For aligned sources we use pink and for non-aligned sources we use teal. The

Triple red geysers have extended jets and their location in Figure 2 suggests they are later pink path phases and thus FRI jets. Namrata et al 2021 infer the jet and wind directions for source 1-289864 to be aligned although this may be difficult to interpret due to its complex structure. We show LOFAR images of the Triple red geysers in Figures 3,4, and 5.  Using triangles we plot the Double red geysers and with rhombuses with plot the Irregular red geysers. These latter sources all appear to have jets and winds that are aligned. Overall, the conclusion appears to be that as we move rightward, alignment begins to be less dominant, which is the model-prescribed behavior.

From Figures 3,4, and 5, we attempt to infer radio morphology for the Triple sources. We classify 1-378770 as an FRI and 1-595166 as an FRII, while 1-289864 remains un-classified due to its complex structure but could conceivably be an FRI. The values of SFR and SM are listed in Table 1. Compatibility with the model suggests that 1-595266 is either on a green path but such that the SFR enhancement is not as great as in similar sources, or that it may have already transitioned through zero black hole spin and the FRII jet is a relic. Dating the wind in this source would, therefore, be particularly useful. Source 1-289864 cannot be classified but we can predict that its deep location in the SFR suppression region of Figure 2 makes it an FRI.

| MaNGA ID | Log (SFR) | Log (SM) |
| --- | --- | --- |
| 1-378770 | -1.33 | 11.46 |
| 1-595166 | -1.04 | 11.06 |
| 1-289864 | -2.44 | 11.06 |

Table 1: The SFR and SM values for the Triple red geysers.

In future work, we plan to explore the cluster richness of these objects, especially when future surveys reveal greater numbers of compact and extended radio sources. The entire FRII phase of a radio source is predicted to be associated with counterrotation, which is a phase during which the ionized disk winds and the black hole jet tend to be aligned. Any compact jets with these features are modeled as the result of low black hole spin which are either approaching zero spin in counterrotation or have experienced a zero-spin transition in the recent past. That is the description in terms of black hole spin and disk orientation. For these latter objects, the system transitioned through zero black hole spin, and jet re-orientation may have taken place, which then leads to SFR suppression. This, again, takes the system onto the pink FRI phase shown in Figure 1.  When the spin is above about 0.1 and lower than about 0.2, however, these corotating systems are prescribed to be FR0 radio galaxies (Capetti et al. 2020a; Garofalo & Singh 2019). In fact, these sources will tend to live longer in intermediate richness environments (Capetti et al. 2020b). In short, the degree of alignment among red geysers should on average determine where they appear on the SFR-SM plane. Aligned sources will tend to occupy regions further up and left, with fewer such sources at lower SFR, while sources with a lack of alignment should tend to live further right and down.

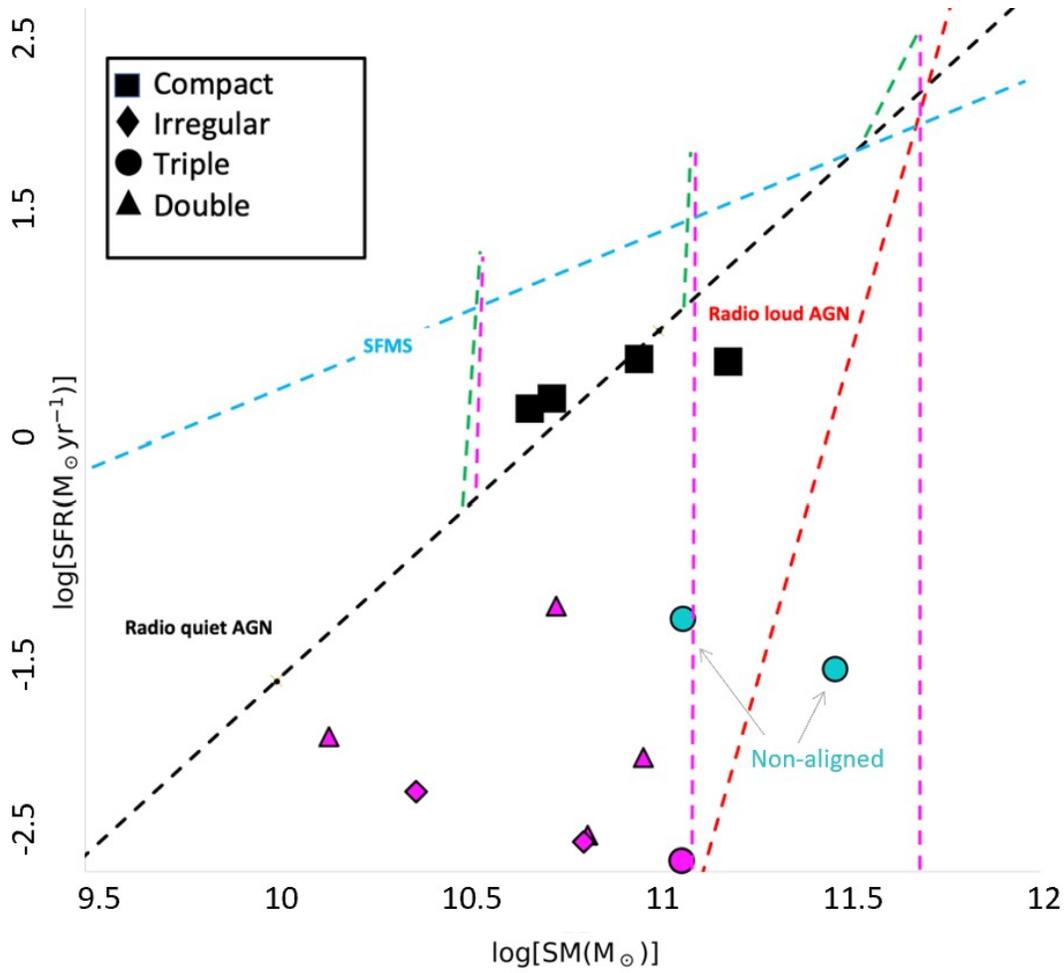

Figure 2: Compact (squares), Triple (circles), Irregular (rhombuses), and Double (triangles) red geysers from Roy et al 2021. Pink is associated with alignment between wind and jet while teal with absence of alignment. The non-aligned sources are also indicated directly. The green and pink paths are the same as in Figure 1.

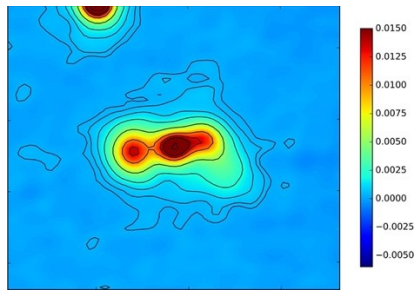

Figure 3: LOFAR image of the source 1-595166. Because the source is classified as an FRII jet (although some disagreement exists about this), it is predicted to be on or close to being on the green paths of Figure 1.

We propose in future work to date the H-alpha winds to estimate their age. As one moves from compact to triple sources, we should observe both an increase in the lack of alignment between jets and winds as well as an increase in the age of the wind. We present these ideas in Table 2. If the radio AGN is currently in a thin or radiatively efficient disk configuration, it produces a disk wind, and therefore the time since the wind dies off is zero. Once the thin disk is replaced by an ADAF, the radiatively driven disk wind ceases to be generated and one can determine the time since that occurrence. This is indicated in the 4$^{th}$ column of Table 2. Although we refer to the angles as the wind and jet angles, strictly speaking this refers to the direction that is normal to the disk surface and to the black hole angular momentum, respectively. If the transition through zero black hole spin is associated with a re-orientation of the disk, the direction of the previous disk normal is now different from the direction of the new disk normal, which is also the direction of the new jet. Hence, the disk and jet are not aligned. This is indicated in the 5$^{th}$ column of Table 2.

| Radio morphology | Excitation class | Environment | Time since wind | Wind & jet angle |
|---|---|---|---|---|
| FRII | High | Isolated | 0 | $0^0$ aligned |
| FRII | Low | Intermediate | $10^8$ years | $0^0$ aligned |
| FR0 | High | Isolated | $5 \times 10^6$ years | $0 - 90^0$ offset |
| FR0 | Low | Intermediate | $5 \times 10^8$ years | $0 - 90^0$ offset |
| FRI | High | Isolated | $15 \times 10^6$ years | $0 - 90^0$ offset |
| FRI | Low | Dense | $4 \times 10^8 - 10^9$ years | $0 - 90^0$ offset |

Table 2: Radio morphology, excitation class, environment, time since wind shuts down, and angle between the wind and the jet, from theory.

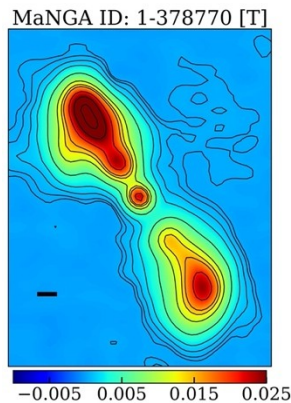

Figure 4: LOFAR image of the source 1-378770. Because the source appears to be of FRI morphology, it is predicted to be a later phase in the evolution of radio galaxies. Hence, with respect to source 1-595166, it should be found at lower SFR, at similar SFR values but further right in Figure 1, or both further right and further down.

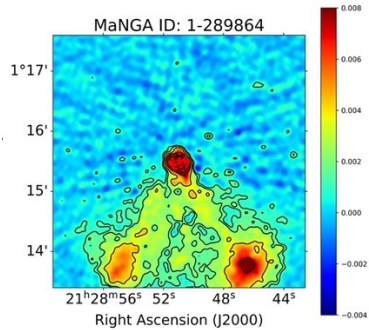

Figure 5: LOFAR image of the source 1-289864. The radio morphology is difficult to establish but its location deep in the SFR suppression region of Figure 2 suggests it is an FRI.

Because counterrotation tends to be triggered in mergers, AGN that result from secular processes will tend to be dominated by corotation between black hole spin and accretion disk. The evolution of such systems will therefore lack the qualitative character of radio mode AGN. The model therefore predicts that AGN in disk galaxies will not experience the powerful SFR suppression associated with FRI jets. Narrow line Seyfert 1s and even jetted Narrow line Seyfert 1s (i.e. Γ NLS1s - Foschini 2010), would tend to evolve as the radio quiet AGN class. If red geysers are indeed different phases in the lifetime of a radio galaxy, the model prescribes them to be located in elliptical galaxies that were triggered in mergers and that they are therefore not located in disk galaxies. It will be important to confirm this as well. While much of this will need to be fleshed out in detail, we simply point out that SFR enhancement/suppression is very much environment dependent which means that jet energy is transferred to the ISM in different ways, and star formation rates, density of stars, and ultimately the M-σ relation should all be impacted. In fact, the different kinds of feedback that we have identified in this work should eventually shed light on all black hole scaling relations.

## 4. Summary and Conclusions

Because the black hole scaling relations relate properties of the stars to the mass of the central black hole, it is reasonable to expect differences in the scaling relations between radio AGN and radio quiet AGN. Feedback from active galactic nuclei has been invoked for decades in order to explain the observed black hole scaling relations. But the evidence has been mixed. Over the last few years, we have identified correlations that explain why some classes of radio galaxies appear to be enhancing star formation rates, while others appear to be suppressing it. But while we have associated these different behaviors with different jet morphology, we had not singled out a well-defined physical mechanism. By focusing on the ubiquitous nature of jet re-orientation in radio galaxies in our model, we extend our exploration of this phenomenon beyond isolated environments for which X-shaped radio galaxies have a preference. In so doing, we identify a more physically robust mechanism for jet-ISM interaction that does not rely on jet morphology. The red-and-dead giant FRI radio galaxies that display strong star formation suppression, according to our picture, are the end product of an initial and short star formation enhancement phase associated with FRII jets. After tens of millions to order hundred million years or so at most, these objects transition into low excitation radio galaxies with corotating black holes whose FRI jets are tilted up to 90 degrees with respect to the wind. Because the wind and jet were aligned, the FRI jet is now tilted with respect to the original jet direction. This allows for a direct coupling between the jet and the ISM, heating it, and eventually shutting down star formation.

This causal relation between jet re-orientation and heat transfer to the ISM we call the Roy Conjecture (from Namrata Roy). In terms of our evolutionary model for radio loud AGN, the number of re-oriented FRI jets should increase where stellar mass is largest and where star formation rates are low. Because disk winds form during near Eddington accretion phases, there will be little evidence of the wind when the star formation suppression has lasted for long periods. Observationally, therefore, one expects to discover lack of alignment between winds and radio jets at times that are not too removed from the transition from counterrotation to corotation. In groups where the evolution of radio loud AGN is faster, one expects to see this phenomenon near where the teal Triple red geysers are located in Figure 2. At very high stellar mass, on the other hand, we predict that we will find such sources closer to log SFR = 0 and log SM = 11.5. While the red geysers provide a statistically small number of sources, their location on the SFR-SM plane is intriguing and future studies of the alignment of jet and winds for objects in this space should be explored. It will be interesting to explore the heating of the ISM for objects that are on opposite sides of the green/pink paths of Figure 1. We predict that we should begin to observe the impact of the FRI jet on the ISM via an enhancement in X-rays. This would constitute the observational link between lack of alignment between wind and jet, and the heating of the ISM that will eventually manifest itself as global SFR suppression. Ultimately, these ideas promise to get us to a deeper understanding of the black hole scaling relations across cosmic time.

Acknowledgment
CBS is supported by the National Natural Science Foundation of China under grant no. 12073021. We thank Namrata Roy for providing us with the red geysers data.

Funding for the Sloan Digital Sky Survey IV has been provided by the Alfred P. Sloan Foundation, the U.S. Department of Energy Office of Science, and the Participating Institutions. SDSS acknowledges support and resources from the Center for High-Performance Computing at the University of Utah. The SDSS web site is www.sdss.org.

SDSS is managed by the Astrophysical Research Consortium for the Participating Institutions of the SDSS Collaboration including the Brazilian Participation Group, the Carnegie Institution for Science, Carnegie Mellon University, Center for Astrophysics | Harvard & Smithsonian (CfA), the Chilean Participation Group, the French Participation Group, Instituto de Astrofísica de Canarias, The Johns Hopkins University, Kavli Institute for the Physics and Mathematics of the Universe (IPMU) / University of Tokyo, the Korean Participation Group, Lawrence Berkeley National Laboratory, Leibniz Institut für Astrophysik Potsdam (AIP), Max-Planck-Institut für Astronomie (MPIA Heidelberg), Max-Planck-Institut für Astrophysik (MPA Garching), Max-Planck-Institut für Extraterrestrische Physik (MPE), National Astronomical Observatories of China, New Mexico State University, New York University, University of Notre Dame, Observatório Nacional / MCTI, The Ohio State University, Pennsylvania State University, Shanghai Astronomical Observatory, United Kingdom Participation Group, Universidad Nacional Autónoma de México, University of Arizona, University of Colorado Boulder, University of Oxford, University of Portsmouth, University of Utah, University of Virginia, University of Washington, University of Wisconsin, Vanderbilt University, and Yale University.